# Mapping quasi-periodic oscillations from the outbursting intermediate polar GK Persei

L. Morales-Rueda[1], M. D. Still[2] and P. Roche[1]

[1] *Astronomy Centre, University of Sussex, Falmer, Brighton BN1 9QH (lmorales@star.maps.susx.ac.uk, pdr@star.maps.susx.ac.uk)*
[2] *Physics and Astronomy, University of St. Andrews, North Haugh, St. Andrews, Fife KY16 9SS (mds1@st-and.ac.uk)*

**ABSTRACT**

We present time-resolved spectrophotometry of GK Per taken on the rise to outburst maximum, in which we detect 5000 s QPOs in the optical emission lines. We use this opportunity to provide a kinematically-resolved analysis of the phenomenon. Observations are consistent with reprocessing off blobs of gas orbiting within the inner accretion disc, and evidence is found for the migration of the QPO source towards the compact star over orbital timescales.

**Key words:**
accretion, accretion discs – binaries: spectroscopic – line: profiles – stars: mass-loss – stars: novae, cataclysmic variables – stars: individual: GK Per.

## 1 INTRODUCTION

In the absence of large magnetic fields, the well-founded model of accretion in close binary stars such as low-mass X-ray binaries (LMXBs) and cataclysmic variables (CVs) involves an accretion disc which transports material from the mass-losing star onto the surface of the compact object (e.g. la Dous 1993). This material is stripped from the donor to form a thin, ballistic gas stream orbiting in the potential well of the accreting object (Lubow & Shu 1975). It has been generally accepted that this stream does not survive the impact with the outer edge of the accretion disc, which is consequently replenished by the new material. Several authors, however, have adopted a scenario where a fraction of the stream can survive this impact and retain its initial ballistic trajectory, before colliding with the inner disc at its point of closest approach to the compact object (Lubow 1989; Hellier 1993). This model has been used to interpret diverse observational properties such as the orbitally-coherent photometric dips in LMXBs (Frank, King & Lasota 1987), white dwarf spin-related modulations from intermediate polars (Hellier 1993) and the velocity fields reconstructed from emission line profiles of nova-like variables (Hellier & Robinson 1994). Encouragingly, SPH calculations of the stream/disc impact by Armitage & Livio (1996) illustrate the ease with which idealized stream material is able to bypass the outer disc edge.

GK Per is an intermediate polar containing a white dwarf with a magnetic field of the order of 1 MG, spinning on a 351 s cycle and accreting from an approximately main sequence donor (Crampton, Cowley & Fisher 1986). Although the magnetic field is suspected to be strong enough to strip the inner accretion disc of material before it reaches the equatorial surface of the white dwarf, forming an accretion curtain, the long 2 d period of GK Per, and consequent wide stellar separation, ensures that a disc should be present. Observational evidence for this is supplied by the recurring dwarf nova outbursts, modelled by Kim, Wheeler & Mineshige (1992) as thermal instabilities within an accretion disc. Time-series analysis of X-ray data obtained during such an outburst by Watson, King & Osborne (1985; hereafter WKO) revealed the presence of $\sim 5000$ s quasi-periodic oscillations (QPOs) – a frequency seemingly unrelated to the white dwarf spin and orbital cycles. The authors suggested that these QPOs were the consequence of modulated mass transfer through the magnetically-channeled accretion curtain, where the 5000 s period is the beat between the spin frequency and material orbiting at the inner edge of the disc. This possibility is strengthened by the identification of 380 s optical QPOs by Patterson (1981). However a reanalysis of the data by Hellier & Livio (1994; hereafter HL) illustrated that the X-ray hardness ratio was anti-correlated with intensity, more suggestive of periodic photoelectric absorption of the X-ray source by cold gas. The relevance of these results to the overflowing accretion stream problem is the frequency of oscillation. Adopting the binary solution of Crampton, Cowley & Fisher (1986) and the limit on disc size of Kim, Wheeler & Mineshige (1992), azimuthally-dependent vertical structure at the outer disc edge has a Keplerian period of $> 1.7 \times 10^4$ s. However, the Keplerian period at the annulus where a disc overflow is expected to impact the inner disc is $4.8 \times 10^3$ s, consistent with observation.

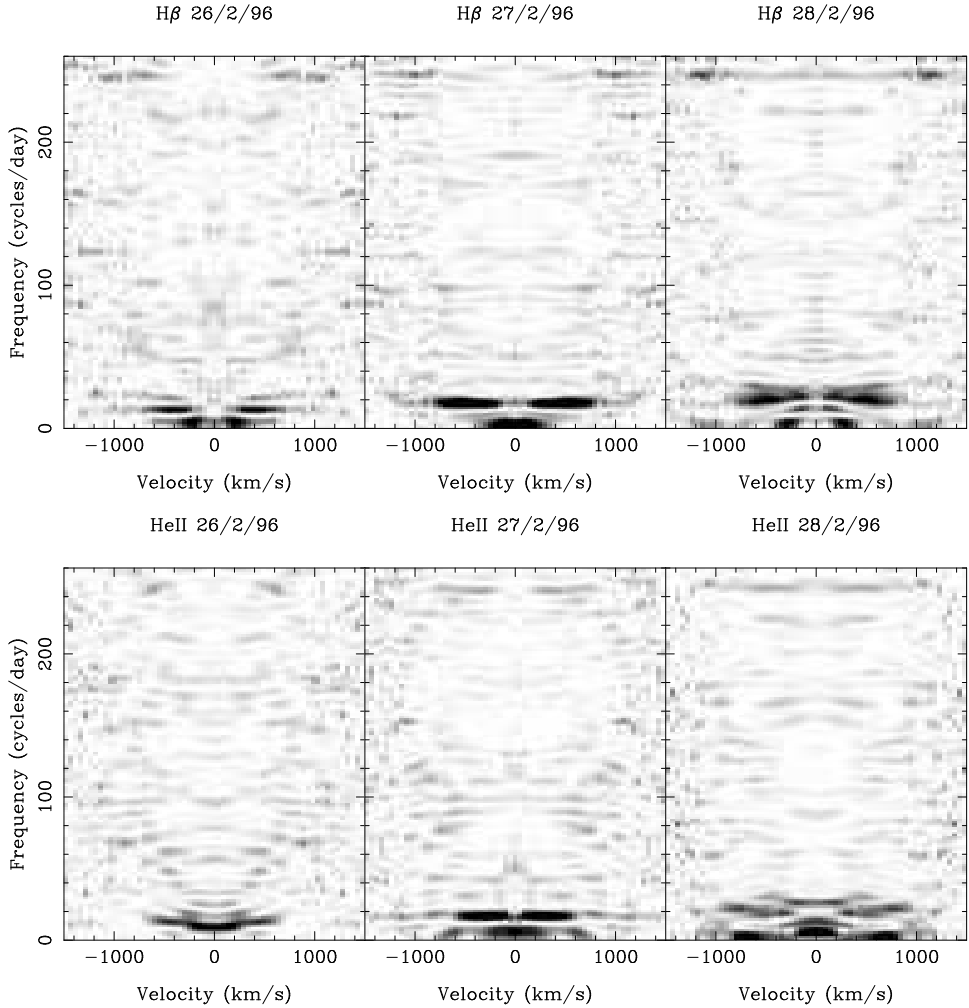

**Figure 1.** Power maps of frequency versus line velocity from the Hβ and He II λ4686 Å emission lines of GK Per, displaying significant power at the 5000 s QPO and spin frequencies.

Therefore the QPO model of HL consists of inhomogeneous 'blobs' of accreting gas joining the inner disc via an overflowing gas stream, which subsequently occult the X-ray source periodically on Keplerian time-scales. The obvious prediction of this model is that QPOs occur at Keplerian velocities appropriate to the inner disc impact point, in this case $V_{\rm Kep} \geq 540 \sin i$ km s$^{-1}$, where $i$ is the orbital inclination. The inclination of GK Per is poorly constrained but from the lack of eclipses $\sin i$ is considered to be between 0.70–0.96, Reinsch (1994). The detection by Reinsch (1994) of power at 2000–7000 s in optical observations during quiescence suggested that the above prediction could be testable.

In this letter we report on optical spectrophotometry of GK Per obtained during the 1996 outburst. We detect the ~5000 s QPOs in the emission lines of this object and show that the velocity dependence of oscillations is generally consistent with the HL overflow model, whereas photometric considerations argue against the variable mass transfer model of WKO.

## 2  OPTICAL QPOS FROM GK PER

The 1996 outburst of GK Per is reported to have begun on February 19 by Mattei (1996). Ishida et al. (1996) report the detection of pulsations of the order of a few thousand seconds in X-ray observations obtained on March 4. Our time-resolved optical spectrophotometry, covering the emission lines of Hβ and He II λ4686 Å, were obtained with the 2.5 m Isaac Newton Telescope, La Palma, on the nights 1996 February 26–28, catching the object at 12th magnitude on the rise to maximum. Exposure times were generally 35 s, with ~70 s dead time, and the spectral resolution corresponds to 80 km s$^{-1}$ at Hβ. Observations were limited by seasonal constraints to 4 h each night, ~3 QPO cycles, providing a total of 342 useable spectra. A full description of the observations, data reduction, and analysis of the outburst phenomena will be provided in a future paper.

In order to search for velocity-resolved periodicities in the data, the emission lines of Hβ and He II λ4686 Å were rebinned onto a constant velocity scale, and the continua were removed from line profiles by subtracting 3-spline fits to line-free regions. Radial velocities for each line were computed using the method of Schneider & Young (1980). This involves convolving a double-Gaussian bandpass with the line profile and determining the minimum of the convolution function. By adopting bandpasses of small widths relative to the line profiles, $\sigma$, and varying discretely the bandpass separation, $s$, it is possible to sample the motion of emission as a function of velocity in the line profile. $\sigma$ was chosen to be 70 km s$^{-1}$ and velocities were sampled by varying $s$ from 0 to 3000 km s$^{-1}$ with discrete steps of 100 km s$^{-1}$. The time series were segregated by date and a search for power over the frequency space was performed using the Lomb-Scargle transform (Scargle 1982). 'Power maps' of frequency versus line velocity are provided in Fig. 1. These maps are mirrored across the rest wavelengths and abscissa units correspond to $s/2$.

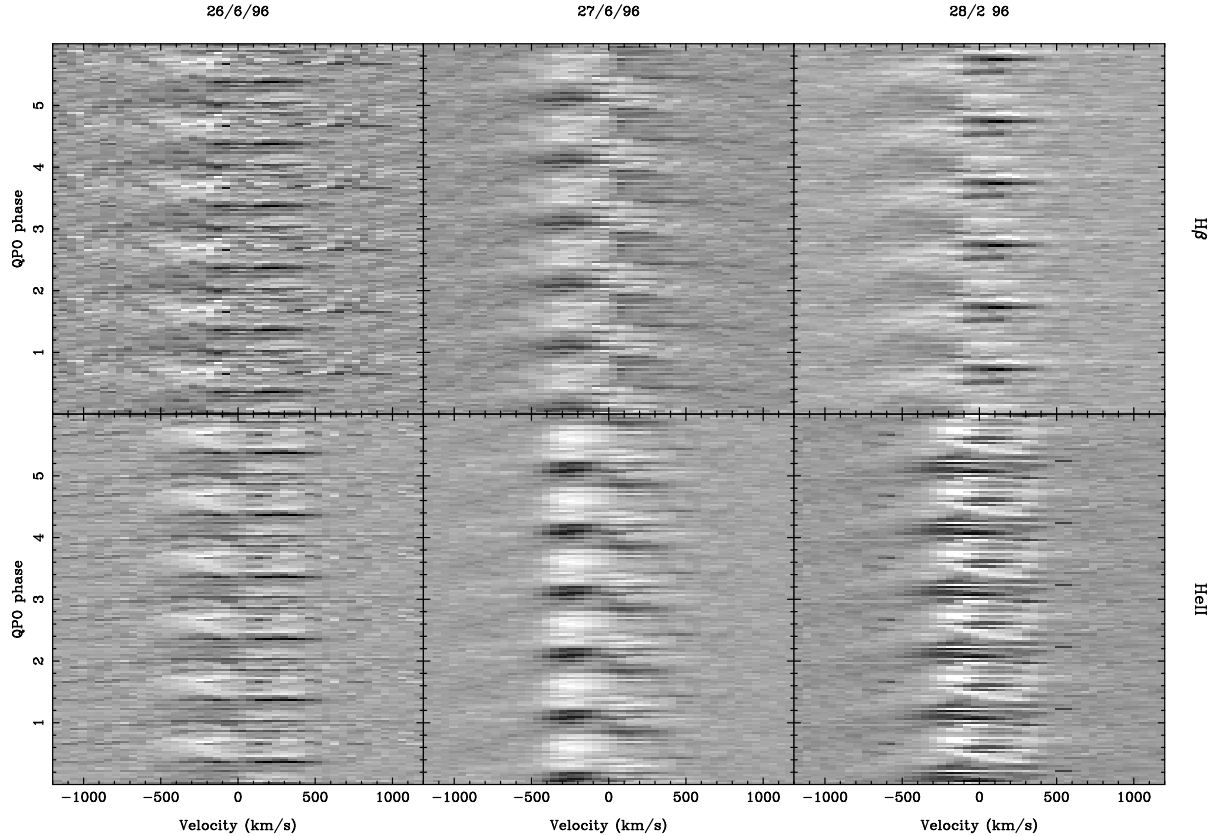

**Figure 2.** The H$\beta$ and He II $\lambda$4686 Å line profile trailed over the QPO cycle determined on 1996 Feb 26–28 at a line-velocity of 400 km s$^{-1}$. Zero-phase corresponds to the first spectrum obtained on each night. Orbital variations due to the white dwarf have been removed and the night-averaged profile subtracted. Grey intensities are on linear, but independent scales, with black corresponding to positive flux residue. Note that the QPO cycle is repeated five times.

We find significant power at three regimes. The first, at $\sim 5$ cycles d$^{-1}$, appears to be a sampling artifact, corresponding to the length of each data set. The second occurs at 246 cycles d$^{-1}$, consistent with the white dwarf spin frequency. The majority of power in this case occurs at the large velocities expected from reprocessed pulsations in the inner-most disc or accretion curtain. The behaviour of these pulsations will be pursued in a future paper. However, of most interest is the detection of power at $\sim 18$ cycles d$^{-1}$, identifiable with the 5000 s QPOs found in X-ray data. More significantly, the maximum characteristic velocities of this power are consistent with $V_{\rm Kep}$ at the impact radius. We also note that although the peak and velocity distribution of the power is consistent between the two lines, the QPOs are evolving from low velocities and low frequencies to higher values over the three nights. Frequency variations were also noted in the X-ray observations by HL. Power maps corresponding to the third night show a more complicated structure in the 18 cycles d$^{-1}$ range.

**Table 1.** The QPO periods from the lines of H$\beta$ and He II $\lambda$4686 Å on the three nights. All periods were found to be significant to 3 sigma within the quoted errors.

| 1996 Feb | $P_{\rm QPO}$ (s) at 400 km/s | $P_{\rm QPO}$ (s) at 800 km/s |
|---|---|---|
| H$\beta$: 26 | 6400 ± 50 | — |
| H$\beta$: 27 | 4909 ± 30 | 4645 ± 25 |
| H$\beta$: 28 | 4214 ± 21 | 4500 ± 25 |
| HeII: 26 | 6595 ± 51 | — |
| HeII: 27 | 5053 ± 30 | 4645 ± 25 |
| HeII: 28 | 4500 ± 25 | 3823 ± 20 |

In order to provide a more qualitative picture of 5000 s variations in the line profiles, the white dwarf motion was removed from the data by shifting the profiles according to the orbital fit of Crampton, Cowley & Fisher (1986), the average profile from each night was subtracted, and the residuals were subsequently folded on the QPO cycle, binned into 30 equally-spaced intervals. The resultant profiles are plotted in Fig. 2 and consist of a positive and negative component shifting in anti-phase across the white dwarf rest velocity. This is indicative of a *single* emission or absorption component in the original untreated profiles. Modulations on the first night are just above the noise level. The first column of Table 1 provides the QPO periods for each map.

## 3   DISCUSSION

We have detected optical 5000 s QPOs in the intermediate polar GK Per on the rise to outburst and resolved their velocity distributions. These are of special interest to investigators who model QPOs as dense blobs of material orbiting the white dwarf within the inner accretion disc. The observations here are consistent with this picture, where the characteristic QPO velocities extend into the inner disc where an overflow impact is likely to occur. However it is equally obvious that QPO power is not restricted to high velocity material and occurs over a range of values down to the order of velocities expected from the outer disc rim, $< 350 \sin i$ km s$^{-1}$. It therefore seems unlikely that we are observing blobs directly unless blob emission is dominated by a broadening mechanism other than the Doppler effect. We can also interpret the optical modulation as variable irradiation of the entire disc from the characteristic QPO radius outwards, caused by the obscuration of the X-ray source by an accreting blob.

Fig. 1 and Table 1 illustrate that the QPO frequency is an approximately constant function of velocity. If line broadening is purely Doppler in nature, this is at odds with the QPO models associated directly with radial or vertical perturbations across the accretion disc (Carroll et al. 1985), which predict that the frequency of oscillation is a linear function of the Keplerian frequency at a given disc annulus. In such a case we would expect the QPO frequency to increase as $V_{\text{Kep}}^3$.

The daily variation of QPO properties may be because we are observing different blobs on each night. However, the evolution of QPO frequencies and velocities to larger values is a natural phenomenon in this model, since if blobs can survive within the disc for durations greater than the dynamical timescale, they are expected to migrate towards the accretor via viscous interaction with the surrounding medium. The QPO distribution on the 3rd night appears more complex. Perhaps this is caused by the deposition of further blobs into the disc, or the fragmentation of the original blob.

A similar time-series analysis to what has been described in Sec. 2 was performed on the data, but sampling the line flux variations as a function of velocity. In these cases power was predictably found on the 5000 s QPOs in discrete bins. However, within the limits of the data, flux is conserved across the integrated line profiles over the QPO cycle. This is consistent with the HL model, provided we have unrestricted viewing over the entire surface of the illuminated disc – a possibility given the predicted range of orbital inclinations. In this case, the WKO model of variable mass transfer through the accretion curtain suffers because an increase in accretion rate is expected to have an observable effect on the line fluxes. However, a reservation concerning the HL model is provided by the result that QPO power determined from the line intensities is biased towards the blue wing of both lines. This is directly observable in the trailed profiles of Fig. 2 where the modulations are stronger in the blue, relative to the red velocities. Because of the persistence of this bias on all three nights, we have difficulty accommodating it in the overflow picture.

In summary, we have found that the kinematically-resolved behaviour of the 5000 s QPO is generally consistent with the disc overflow scenarios discussed by Lubow (1989), HL and Armitage & Livio (1996). We favour a model where vertically-extended blobs, fixed on dynamical timescales within the inner accretion disc, provide modulated reprocessing through obscuration of, or illumination by, the central X-ray source. We also find evidence for the evolution of this blob towards the compact object. Some reservations are provided by the observation that QPO power is biased towards approaching material in a frame rotating with the white dwarf orbit.

## ACKNOWLEDGMENTS

We thank Tom Marsh for making available his reduction software and the Nuffield Foundation for a grant to PR in order to facilitate this collaborative work. The Isaac Newton Group of telescopes are operated on the island of La Palma by the Royal Greenwich Observatory in the Spanish Observatorio del Roque de los Muchachos of the Instituto de Astrofísica de Canarias.

This paper has been produced using the Blackwell Scientific Publications LaTeX style file.